\documentclass[aps,pre,twocolumn,groupedaddress]{revtex4-1}
\usepackage{graphicx}
\usepackage{enumerate}
\usepackage{epstopdf}
\usepackage{amsmath, amssymb, upgreek, textcomp}

\begin{document}

\title{Extended phase space description of human-controlled systems dynamics}
 
\author{Arkady Zgonnikov}
\email[]{arkady.zgonnikov@gmail.com}

\author{Ihor Lubashevsky}
\email[]{i-lubash@u-aizu.ac.jp}

\affiliation{University of Aizu, Tsuruga, Ikki-machi, Aizuwakamatsu, Fukushima 965-8580, Japan} 

\begin{abstract}
Humans are often incapable of precisely identifying and implementing the desired control strategy in controlling unstable dynamical systems. That is, the operator of a dynamical system treats the current control effort as acceptable even if it deviates slightly from the desired value, and starts correcting the actions only when the deviation has become evident. We argue that the standard Newtonian approach does not allow to model such behavior. Instead, the physical phase space of a controlled system should be extended with an independent phase variable characterizing the operator motivated actions. The proposed approach is illustrated via a simple non-Newtonian model capturing the operators fuzzy perception of their own actions. The properties of the model are investigated analytically and numerically; the results confirm that the extended phase space may aid in capturing the intricate dynamical properties of human-controlled systems.
\end{abstract}


\maketitle

\section{Introduction}
The basic physical notions and mathematical formalism have been successfully employed in modeling social and psychological phenomena. The notions of Newtonian mechanics were used in social force models for traffic dynamics and crowd behavior~\cite{helbing2001traffic},~\cite{helbing2000simulating}. The statistical physics framework, namely, the master equation approach, has been widely used in describing the opinion formation and language evolution~\cite{castellano2009statistical}. Nonetheless, despite the gained success in describing social phenomena in mathematical terms, up to now there is a strong demand for notions and models reflecting the unique properties of human beings~\cite{vallacher2009applications}. Such features as feelings, emotions, intentions, and beliefs distinguish humans from inanimate objects studied in physics. Development of specific notions and formalism capturing the peculiarities of human behavior at the level of individuals can enable us to model, simulate and better understand 
complex 
social phenomena met in everyday life.

One of the cornerstones of modern physics widely met in social psychology~\cite{nowak1998dynamical} is the notion of fixed-point attractor, or equilibrium. For instance, a person achieving and maintaining a certain end-state or goal can be formally treated as a dynamical system drifting towards an equilibrium point in the corresponding phase space~\cite{carver2002control}. Likewise, one may  consider an entity controlled directly by a human operator whose purpose is to maintain its stability. In this case the system dynamics as a whole also can be described by a fixed-point attractor. External or internal factors may cause the system to deviate from the equilibrium, but if the operator is capable of handling such perturbations, the system will eventually evolve to the desired state.

Recent advances in the field of human control give evidence to the fact that humans do not generally operate the systems under their control in a precise way. Maintaining the system exactly at the desired position requires the ability of the operator to keep perfect awareness and to react immediately even to the smallest deviations. Meanwhile, experimental studies have revealed that the considerable response latency and the effects of noise in the sensorimotor system prevent human operators from implementing continuous control strategies (see, e.g.,~\cite{loram2011human} and references therein). Instead, the discontinuous, or intermittent control~\cite{cabrera2002onoff} is found to be efficient in the presence of time delays and random perturbations in human-controlled processes. The ``drift and act'' pattern of human control has been detected, e.g., in aircraft landing~\cite{gestwa2011using}, stick balancing at the fingertip~\cite{milton2009balancing} and postural control during quiet standing~\cite{
loram2005human}.

In each of these processes human operator prefers to ignore small deviations of the dynamical system from the desired state, starting the active control over the system only when the deviation becomes too large to ignore. The reasons for such behavior vary depending on the properties of the particular system. For instance, while controlling the systems that are relatively sensitive to human response (e.g., balancing a stick), operators ignore small deviations in order not to destabilize the system by the imprecise corrective actions~\cite{loram2011human}. In the processes with relatively slow dynamics (e.g., car following) human operators tend to ``satisfice'' rather than to optimize~\cite{boer1999car}: an operator stays relaxed while the current situation is acceptable, taking the control over the system only when she is uncomfortable with the deviation from the desired state. 

\begin{figure}
\centering
 \includegraphics[width=1.0\columnwidth]{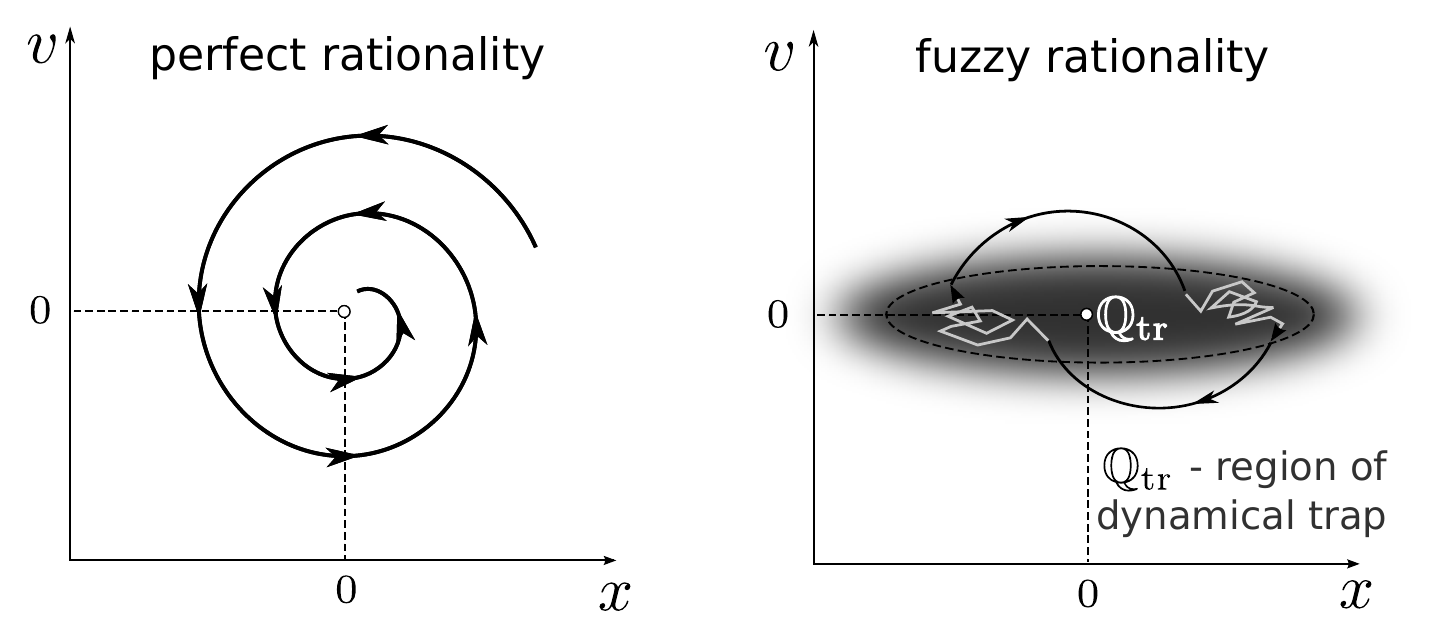}
\caption{The phase space structure of the dynamical system under human control in case of perfect and fuzzy operator rationality. Figure adapted from~\cite{lubashevsky2012dynamical}.}
\label{fig:DT}
\end{figure}

The aforesaid allows us to conclude that human operators, at least in some situations, do not distinguish between the optimal state of a controlled system and sub-optimal states from its vicinity. This property of human behavior is a manifestation of the human fuzzy rationality~\cite{dompere2009fuzzy}. In the general case the operator fuzzy rationality makes the standard equilibrium point formalism not applicable for describing the dynamics of human-controlled systems. Instead, such systems exhibit some complex time-dependent patterns of behavior near the virtual equilibrium~\cite{zgonnikov2012computer}. Up to now there have been a few attempts to develop a mathematical formalism capturing the effects of human fuzzy rationality. In particular, the model of fixed human reaction threshold is commonly used in applied studies~\cite{asai2009model},\cite{gestwa2011using},\cite{milton2009balancing},\cite{milton2009discontinuous}, but still there is a lot of uncertainty about the intrinsic mechanisms causing the 
anomalous behavior of 
the systems under human control (see, e.g.,~\cite{cabrera2012stick}). The dynamical trap model~\cite{lubashevsky2012dynamical},\cite{lubashevsky2003noised},\cite{lubashevsky2002long}, being a certain version of the fuzzy threshold concept, is another alternative to the standard fixed-point attractor in complex sociopsychological systems. Both of the two models capture the fuzziness of the desired end-state by introducing a certain region around the virtual equilibrium, where each state is treated as acceptable by the operator (Fig.~\ref{fig:DT}, right frame).
However, these models consider the operator behavior to be strictly optimal outside the region of acceptable states.
It means that in driving the system towards this region the operator generally follows some predefined control law. Let us, for example, consider a physical system whose dynamics is specified by its coordinate $x$ and velocity $v=dx/dt$. Then the actions of an operator can be represented as a certain function $a=a(x,v)$ of the phase variables $\{x,v\}$, maybe with some time delay. As a result, the behavior of such a system governed by the operator actions is completely described by a differential equation similar to $\dot{v} = F(x,v,a)$.

\begin{figure}
\centering
 \includegraphics[width=0.8\columnwidth]{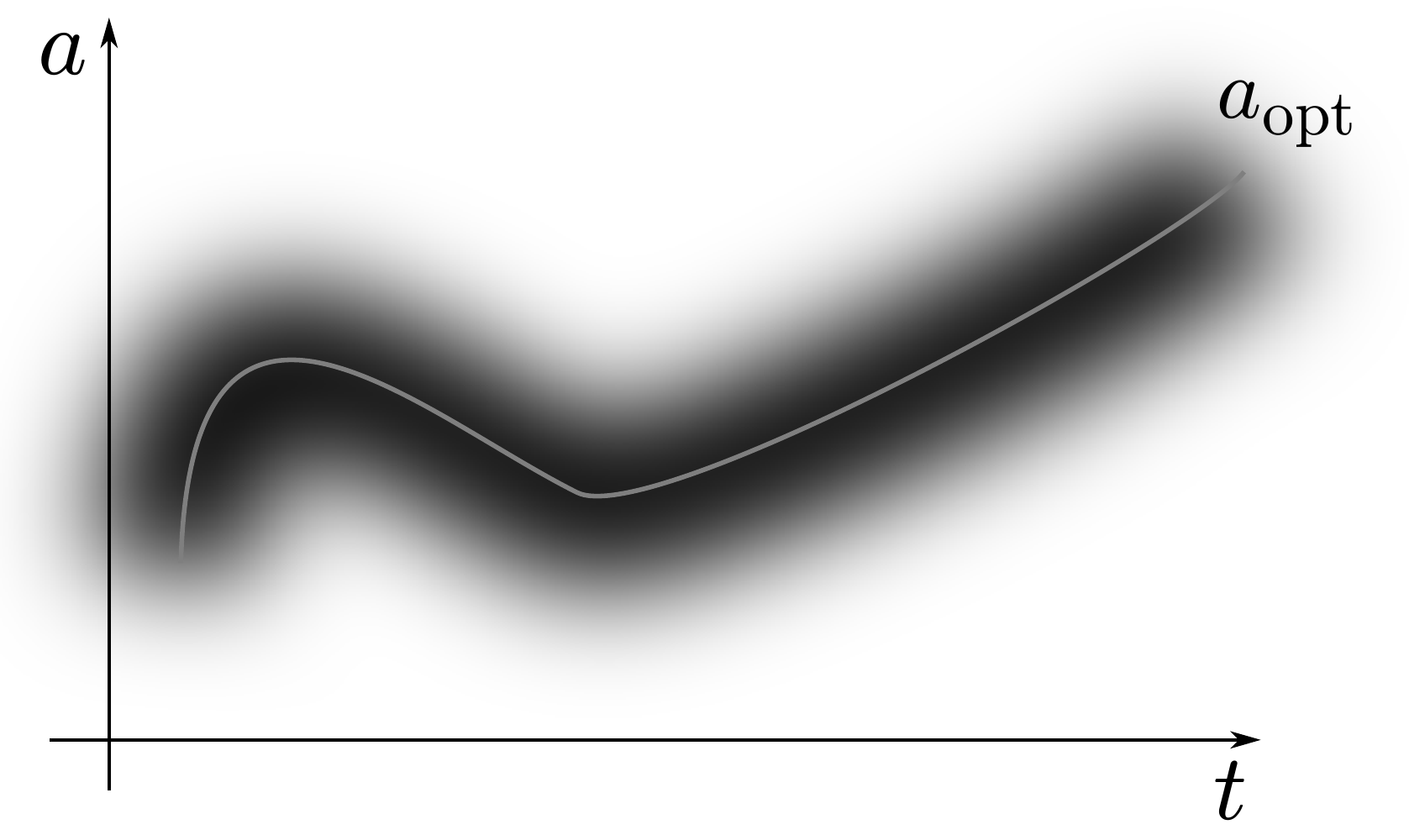}
\caption{Action dynamical trap: the vicinity of the optimal strategy $a_{\textrm{opt}}(t)$ in the space of all action strategies $\{a(t)\}$.}
\label{fig:ADT}
\end{figure}

Still, one may claim that humans are often unable to implement the desired control strategy $a(x,v)$ precisely. Rather, the operator is only able to detect if the current control effort is worth changing when the deviation from the desired strategy becomes large enough. In other words, at each instant there is a whole fuzzy set of control parameter values acceptable for the operator (Fig.~\ref{fig:ADT}). The common approach to modeling this effect is to introduce an additive noise term (see, e.g.,~\cite{treiber2006understanding}), that is justified in some situations. However, this approach does not reflect some characteristic features of human control (e.g., the on-off intermittency~\cite{cabrera2002onoff}). Besides, a wide class of intricate phenomena observed in the systems of interacting individuals still remains unexplained.

In the present paper we argue that the effect of human fuzzy rationality extends beyond the concept of desired end-state to the notion of action strategy. Appealing to the dynamical traps framework we try to capture the fluctuations of the operator actions in the vicinity of the virtual optimal control strategy. We extend the two-dimensional system phase space $(x,v)$ with acceleration $a$ as an independent phase variable that is perceived, evaluated, and indirectly controlled by the operator. By cost of introducing the non-Newtonian variable $a$ we gain the possibility of describing mathematically the fuzzy set of acceptable suboptimal action strategies which is further referred to as the action dynamical trap.

\section{Model background}
Hereafter we explain the idea of the basic dynamical trap model by employing a simple example of a noisy human-control\-led dynamical system with no internal dynamics. The phase space of the system comprises the coordinate $x$ and the velocity $v$. The goal of the operator is to maintain the system at the desired state, the origin, by implementing the optimal in some sense control strategy $a_{\textrm{opt}}(x,v)$. However, if the system currently resides in some vicinity of the desired state, the operator prefers to halt active control over the system. The equations describing the system dynamics under the operator control are written as follows 
\begin{equation}
\label{eq:general}
 \begin{aligned}
  \dot x &= v, \\
  \dot v &= \Omega(x,v)a_{\textrm{opt}}(x,v) + \varepsilon f(t),
\end{aligned}
\end{equation}
where $f(t)$ is the random force with the amplitude $\varepsilon~\ll~1$. The cofactor $\Omega(x,v)$ is some function such that $\Omega(x,v) \approx 1$ for all the values $(x,v)$ that are far enough from the origin and $\Omega(x,v) \ll 1$ in a certain neighborhood $\mathbb{Q}_{\textrm{tr}}$ of the origin (Fig.~\ref{fig:DT}, right frame).

In order to explain the meaning of the cofactor $\Omega(x,v)$ we consider the behavior of the operator approaching the desired state $(x=0,v=0)$. When the current state is far from the origin, the operator perfectly follows certain optimal action strategy $a_{\textrm{opt}}(x,v)$. If the current position is recognized as ``good enough'', i.e.,  $(x,y)\in \mathbb{Q}_{\textrm{tr}}$, the operator halts active control over the system. So, during a considerable period of time the system is affected only by random factors of a small amplitude; in other words, the system is ``trapped'' in a vicinity of the desired position. Therefore, $\mathbb{Q}_{\textrm{tr}}$ is called the area of dynamical trap. One may notice that in the case of linear feedback strategy $$a_{\textrm{opt}} \propto - (x + \sigma v),$$ where $\sigma > 0$ is a constant damping parameter, the given system under human control is analogous to the physical system of a damped harmonic oscillator. This allows us to call system~\eqref{eq:general} the 
oscillator with dynamical trap.

The oscillator with dynamical trap captures the basic behavior properties of the fuzzy rational operator, i.e., the operator who does not react to small deviations from the desired phase space position. When the system deviates significantly from the goal state, the operator decides to start controlling the system in order to return it to an acceptable state. This can be achieved by varying the control parameter, namely, the acceleration, in a way that is optimal in some sense.

Let us appeal to the car following, which is a characteristic example of a dynamical system governed by operator with fuzzy rationality~\cite{lubashevsky2003bounded}. Car drivers are unable to continuously keep perfect awareness of the surrounding situation, so they usually set the acceleration to some constant value based on the current circumstances. Once fixed, the value of acceleration is changed only when the driver realizes that the deviation from some ``optimal'' acceleration value has become too large to be ignored. In other words, considerable deviations of the current acceleration $a$ from the optimal value $a_{\textrm{opt}}$ cause the operator to start active control over the car motion. However, when the difference $a-a_{\textrm{opt}}$ is rather small, there are no stimuli for the driver to act, i.e., to change the acceleration. Thus, one may imagine a certain region around the optimal strategy $a_{\textrm{opt}}(x,v)$, wherein each strategy is regarded as acceptable (Fig.~\ref{fig:ADT}). Instead 
of precisely following the optimal strategy, the operator just keeps the actually implemented strategy inside this region, making some corrections only when the mismatch $a-a_{\textrm{opt}}$ exceeds some fuzzy threshold. For this reason the region of acceptable strategies around $a_{\textrm{opt}}$ will be called the action dynamical trap. The ``thickness'' of the action dynamical trap is determined by the capacity of the operator perception and levels of concentration and motivation to follow the optimal control strategy. The action dynamical trap model is proposed to capture the discussed effects of fuzzy rationality in choosing and implementing the action strategies in human-controlled dynamical processes.

\section{Action dynamical trap model}
We start our speculations from considering the original dynamical trap model described by equations~\eqref{eq:general}. In order to elucidate the basic properties of the model, we exclude the random factors from the scope of the present paper, i.e., we consider $f(t)\equiv 0$. 

The pivot point of the proposed approach is that we regard human actions as an independent component of the system rather than some predetermined function of its physical state. We extend the physical phase space $\{x,v\}$ by introducing a new phase variable, in the given case, the system acceleration $a$, i.e.,
\begin{equation*}
	\{x,v\}\to \{x,v,a\}.
\end{equation*}
It enables us to ascribe to the system an additional degree of freedom corresponding to the operator actions. Now, the model capturing the dynamical trap effect in controlling the deviation $a-a_\text{opt}$ is written as
\begin{equation}
\begin{aligned}
  \label{eq:model}
  \dot x &= v\,, \\
  \dot v &= a\,, \\
  \tau \dot a &= -\Omega_a\big(a-a_{\textrm{opt}}(x,v)\big)\big(a-a_{\textrm{opt}}(x,v)\big),
\end{aligned}
\end{equation}
where $\tau$ is the operator reaction time parameter, and functions $a_{\textrm{opt}}(x,v)\,\, \textrm{and}\,\, \Omega_a(a-a_{\textrm{opt}})$ are to be specified.

We define the operator control strategy as a linear feedback aimed at maintaining the system at the origin: $a_{\textrm{opt}}(x,v)=-\omega^2(x + \frac{\sigma}{\omega} v)$, where $\omega\,\textrm{and}\,\sigma$ are non-negative constant coefficients. However, as the operator is fuzzy rational, the optimal control strategy should incorporate the dynamical trap effect in correcting the velocity variations:
\begin{equation*}
 a_{\textrm{opt}}(x,v) = -\Omega_{v}(x,v)\omega^2(x+ \frac{\sigma}{\omega} v).
\end{equation*}
Thus, the control strategy $a_\text{opt}$ is optimal from the standpoint of fuzzy rational human operator. 

\begin{figure}
\centering
 \includegraphics[width=0.8\columnwidth]{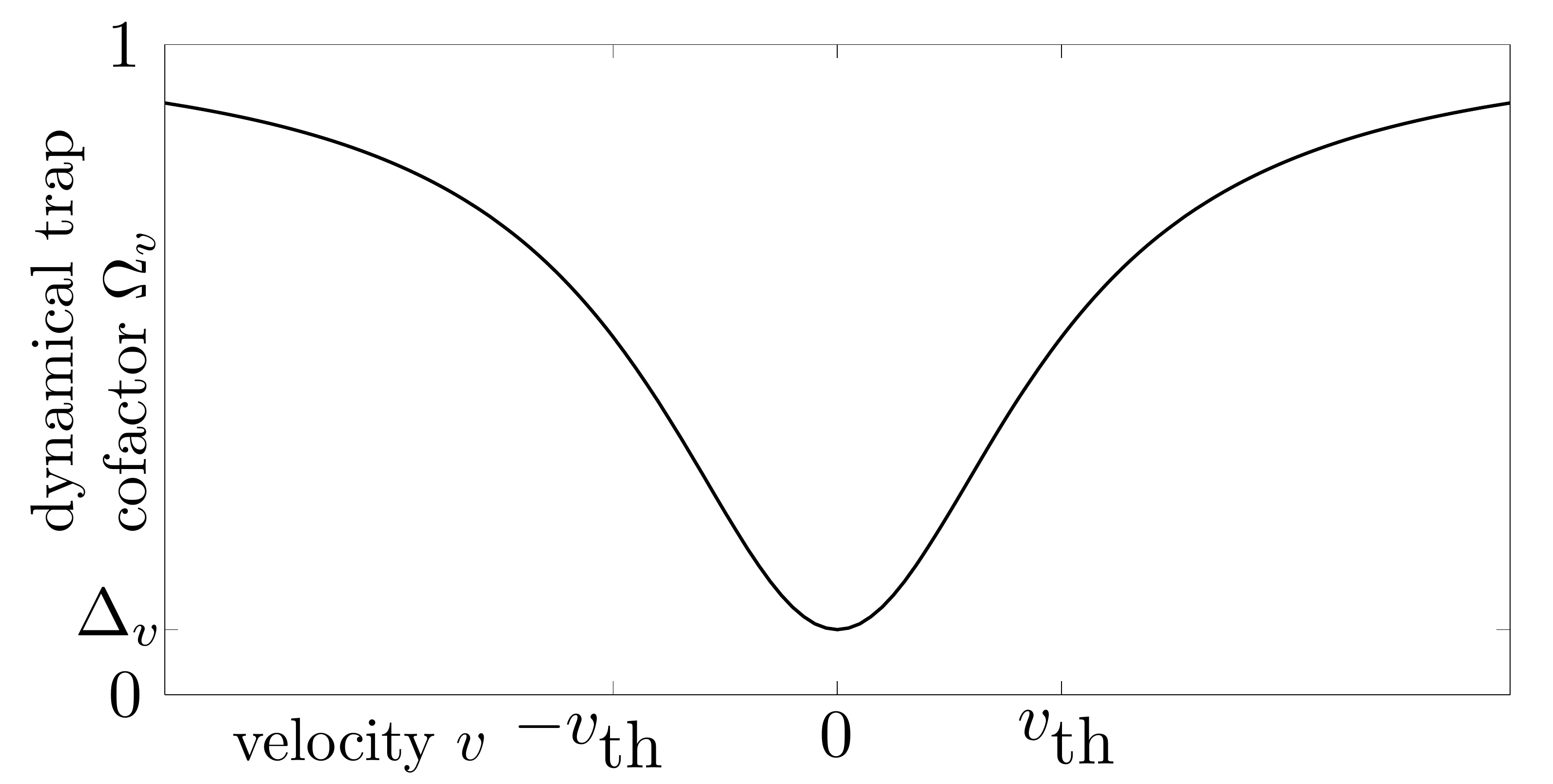}
\caption{The dynamical trap cofactor $\Omega_v(v)$.}
\label{fig:omega}
\end{figure}

Here the dynamical trap cofactor $\Omega_v(x,v)$ is claimed not to depend on $x$. It reflects the assumption that the control over the system velocity $v$ is of prior importance for the operator comparing to the control over the coordinate $x$. The desired effect can be mimicked by any function $\Omega(v)$ such that $\Omega \ll 1$ if $v\approx 0$ and $\Omega \approx 1$ otherwise. Without loss of generality we use the ansatz 
\begin{equation}
\Omega(x,v):=\Omega_v(v)=\frac{\Updelta_{v}v_{\textrm{th}}^2+v^2}{v_{\textrm{th}}^2+v^2}\,,
\label{eq:omega_v}
\end{equation}
where $v_\textrm{th}>0$ is the threshold value of velocity and $\Updelta_v \in [0,1]$ is the dynamical trap intensity coefficient (Fig.~\ref{fig:omega}). When $\Updelta_v$ equals unity, there is no dynamical trap effect \textemdash the operator is strictly rational and reacts even to the tiniest deviations. The case $\Updelta_v=0$ matches the situation when the operator ignores the small deviations but engages actively in the control over the system when the deviation becomes large enough.
One may notice that if we set $\Omega_a(a-a_{\textrm{opt}}\big(x,v)\big)\equiv 1$, system~\eqref{eq:model} describes following the optimal action strategy $a_{\textrm{opt}}$ precisely by the operator whose reaction time is $\tau$. As the human operator is not capable of doing so in the case of small deviations $a-a_\text{opt}$,  we write 
\begin{equation*}
 \Omega_a(a-a_{\textrm{opt}}) = \frac{\Updelta_{a}a_{\textrm{th}}^2+(a_{\textrm{opt}}-a)^2}{a_{\textrm{th}}^2+(a_{\textrm{opt}}-a)^2},
\end{equation*}
where, in analogy to \eqref{eq:omega_v}, $\Updelta_a \in [0,1]$ indicates the presence of the action dynamical trap and $a_\textrm{th}$ is the threshold in perceiving acceleration deviations from the optimal value. 

In order to reduce the number of system parameters we change the time and spatial scales as follows
 \begin{equation*}
  t \to t \frac{1}{\omega} \,,\quad
  x \to x \frac{a_{\textrm{th}}}{\omega^2}\,.
 \end{equation*}
It is easy to check that in these dimensionless units parameters $\omega$ and $a_{\textrm{th}}$ are both equal to unity. Thus, the above expressions for $\Omega_a$ and $a_{\textrm{opt}}$ take form
\begin{equation}
\label{eq:a_opt}
\begin{aligned}
 \Omega_a(a-a_{\textrm{opt}}) &= \frac{\Updelta_{a}+(a_{\textrm{opt}}-a)^2}{1+(a_{\textrm{opt}}-a)^2},\\
 a_{\textrm{opt}}(x,v) &= -\Omega_{v}(v)(x+ \sigma v).
\end{aligned}
\end{equation}

\section{Dynamics of an oscillator with action dynamical trap}
System \eqref{eq:model}--\eqref{eq:a_opt} possesses the only equilibrium point at the origin. Linear stability analysis reveals that this equilibrium is stable for all the values of the system parameters $\sigma$, $\tau$, and $\Updelta_a$ such that 
\begin{equation}\label{eq:stab}
  \frac{\tau}{\sigma} < {\Updelta_a}.
\end{equation}  
If the effect of the action dynamical trap is absent, $\Updelta_a = 1$, the system is stable for $\tau < \sigma$, i.e., when the operator reaction time $\tau$ is relatively small and (or) the capability of suppressing the velocity deviations $\sigma$ is relatively high. 

When the action dynamical trap effect comes into play, $\Updelta_a \ll 1$, system \eqref{eq:model}--\eqref{eq:a_opt} is stable only if $\tau\ll\sigma$. This may be interpreted in such sense that the operator can not precisely maintain the desired state of the system, unless the operator's reaction is almost immediate ($\tau \ll 1$) or the velocity feedback gain $\sigma$ is extremely large. Moreover, when $\Updelta_a$ reaches zero, the system governed by equations \eqref{eq:model}--\eqref{eq:a_opt} becomes unstable at the origin regardless the values of the other parameters. It is notable that the system stability does not depend on the parameters $\Updelta_v$ and $v_{\textrm{th}}$ quantifying the intensity of the velocity dynamical trap and the velocity perception threshold, respectively.

In the present paper we focus on the operator affected by both velocity and acceleration dynamical traps: $\Updelta_v=\Updelta_a=0$. We also comment briefly on the case when the operator is perfectly rational in controlling either velocity ($\Updelta_v=1$) or acceleration ($\Updelta_a=1$) deviations. The intermediate values of the parameters $\Updelta_{a,v}$ far from the boundary values have in fact little physical meaning, corresponding to the hypothetical case when the operator stays focused on controlling the small deviations, but at the same time applies reduced effort in doing so. For this reason, although the below results hold for any $\Updelta_{a,v}$, we refrain from the detailed analysis of the system dynamics in case of $0 < \Updelta_{a,v} < 1$.

We analyzed the behavior of system \eqref{eq:model}--\eqref{eq:a_opt} numerically under the adopted assumptions for various values of system parameters. The absolute and relative error tolerance parameters of the routine used for the numerical simulations were chosen in a way that varying them tenfold could not affect the results of the simulations. The initial conditions for simulations were formed by assigning small random values to the phase variables. 

We observed two major patterns of the system dynamics depending on the parameters $\sigma$, $\tau$ and $v_{\textrm{th}}$. The system either performs periodic oscillations or becomes uncontrollable by the operator, with all phase variables exhibiting unbounded growth.

\begin{figure}
\centering
 \includegraphics[width=1.0\columnwidth]{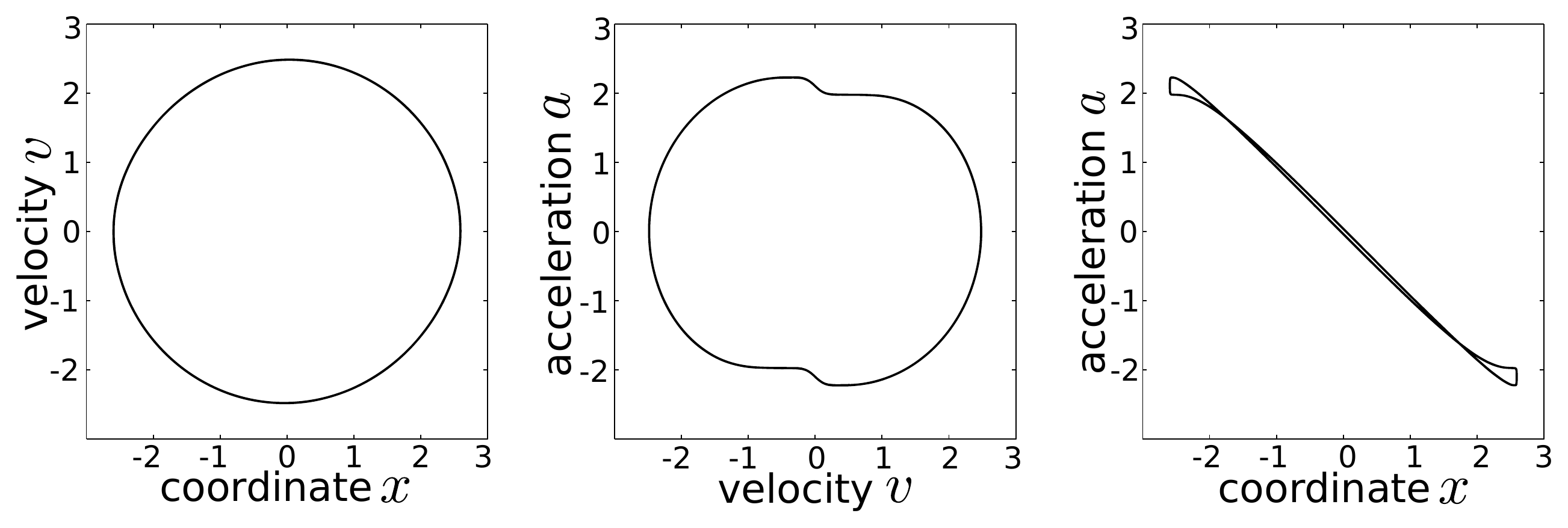}
\caption{The projections of the limit cycle formed by the phase trajectory of system \eqref{eq:model}--\eqref{eq:a_opt}. The values of parameters used for simulation are $\sigma=1,\,\tau=0.9,\,v_{\textrm{th}}=0.2,\,\Updelta_a=\Updelta_v=0$}
\label{fig:phase}
\end{figure}

\begin{figure}
\centering
 \includegraphics[width=1.0\columnwidth]{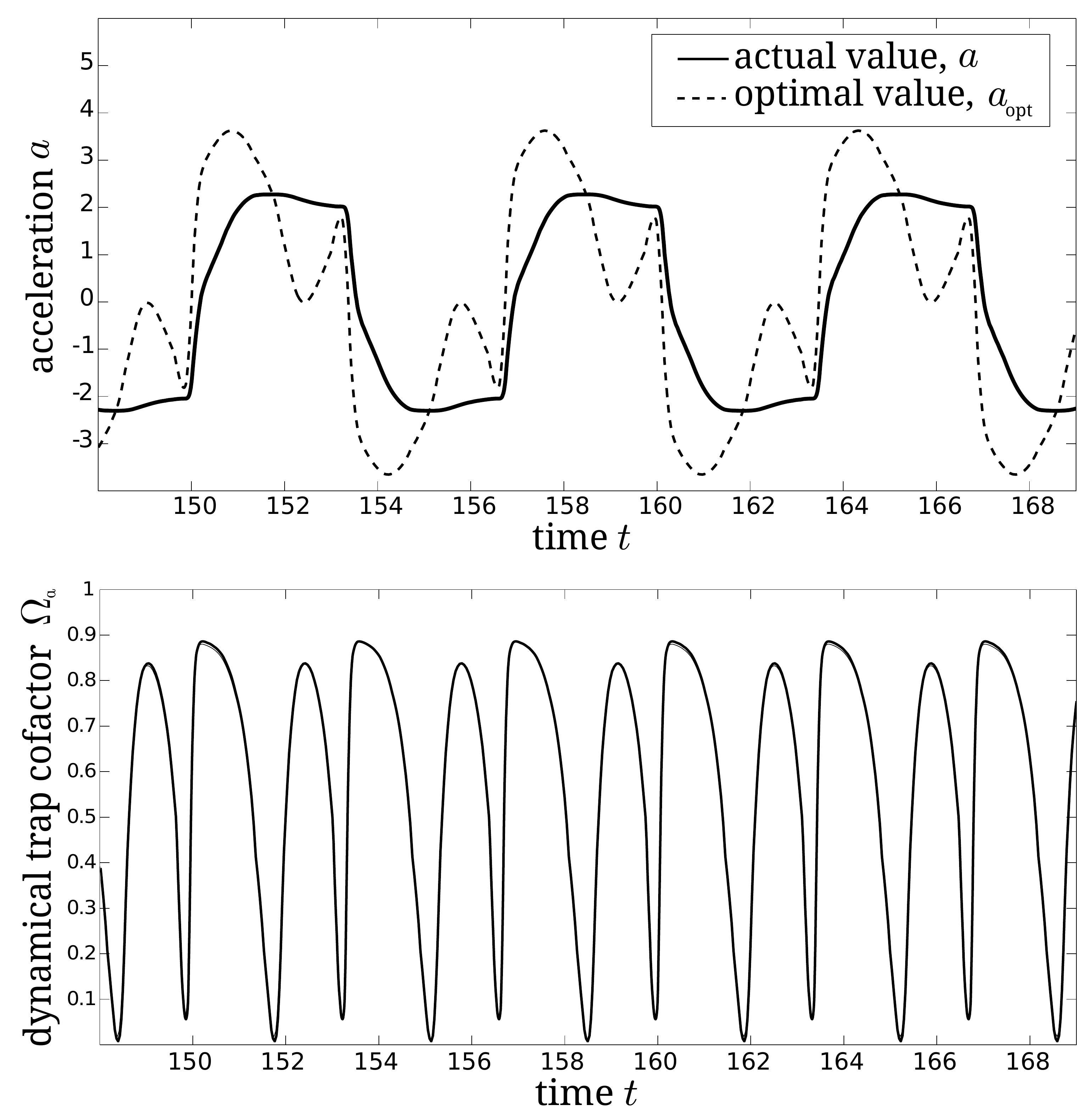}
\caption{Time pattern fragments of the actual and optimal acceleration (top frame) and corresponding dynamical trap cofactor (bottom frame). The fragments match the phase portrait shown in Fig.~\ref{fig:phase}.}
\label{fig:accel}
\end{figure}

Generally, the periodic behavior can be observed when the operator response latency $\tau$ is in some sense small and (or) the feedback gain $\sigma$ is relatively large. The form of the found limit cycle almost does not depend on the particular values of these parameters. Fig.~\ref{fig:phase} represents the example of the limit cycle found for $\sigma=1$, $\tau=0.9$, $v_{\textrm{th}}=0.2$. The fragment of the acceleration time pattern $a(t)$ corresponding to this phase portrait is depicted in the top frame of Fig.~\ref{fig:accel}, as well as is the evolution of the optimal action strategy $a_{\textrm{opt}}(t)$. As clearly seen, the implemented action strategy remains in the vicinity of the optimal one. When the difference between these two strategies becomes sufficiently small, the acceleration growth ratio is also small. It reflects the fact that under this condition the operator almost does not change the control variable, $a$, for a certain period of time. However, when the deviation from the optimal 
action strategy becomes large, the operator behavior turns to be active and the actual acceleration changes fast. This is also reflected in the bottom frame of Fig.~\ref{fig:accel}, where the time pattern of the dynamical trap cofactor $\Omega_a$ is represented. The values of $\Omega_a$ near unity correspond to the periods of the acceleration active growth or decrease, while the stagnation of $a$ is characterized by values of $\Omega_a$ close to zero. When $a-a_{\textrm{opt}}(t)$ becomes large, $\Omega_a$ ``switches on'', and the operator starts to actively control the system. Occasionally only little effort is needed to adjust the current control strategy to the optimal one (see, for instance, $t\approx156$ in Fig.~\ref{fig:accel}), however, sometimes the operator has to correct the actions substantially (e.g., $t\in [160,161]$).

\begin{figure}
\centering
 \includegraphics[width=1.0\columnwidth]{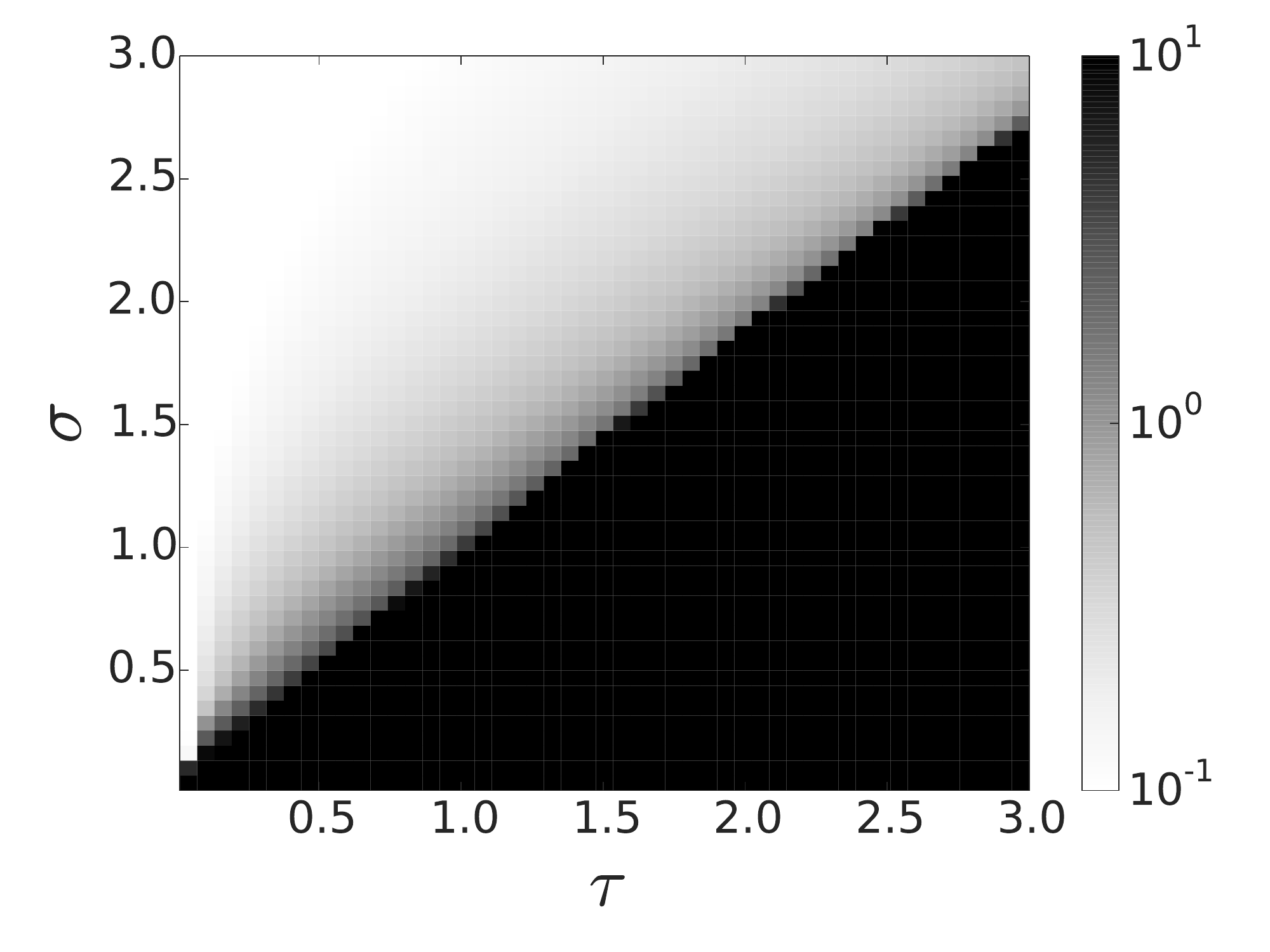}
\caption{Amplitude $v_{\textrm{max}}$ of the system velocity oscillations depending on the parameters $\tau$ and $\sigma$. The brightness of each point is associated with the numerically calculated motion amplitude of the system~\eqref{eq:model}--\eqref{eq:a_opt} for parameters $v_{\textrm{th}}=0.2, \, \Updelta_a=\Updelta_v=0$. The dark grey region corresponds to the unbounded motion of the system.}
\label{fig:ampl}
\end{figure}

As $\tau$ grows and (or) $\sigma$ decreases, the amplitude of the oscillations increases and eventually the periodic pattern evolves to the uncontrolled motion. Fig.~\ref{fig:ampl} demonstrates the dependency of the system velocity amplitude on these two parameters (for some fixed value of $v_{\textrm{th}}$). One can note that for any fixed value of $\tau$ the oscillation magnitude monotonically decays with $\sigma$. Indeed, the better the operator can handle the velocity deviations, the closer the system is to the desired state. Similarly, the limit cycle shrinks as $\tau$ decreases: as the operator reaction becomes faster, it becomes easier to keep the system near the desired position. On the contrary, the operators with large $\tau$ and small $\sigma$ are not capable of controlling the system. Such operators destabilize the system by unintelligent actions, so the phase variables reach infinite values.

\begin{figure}
\centering
 \includegraphics[width=1.0\columnwidth]{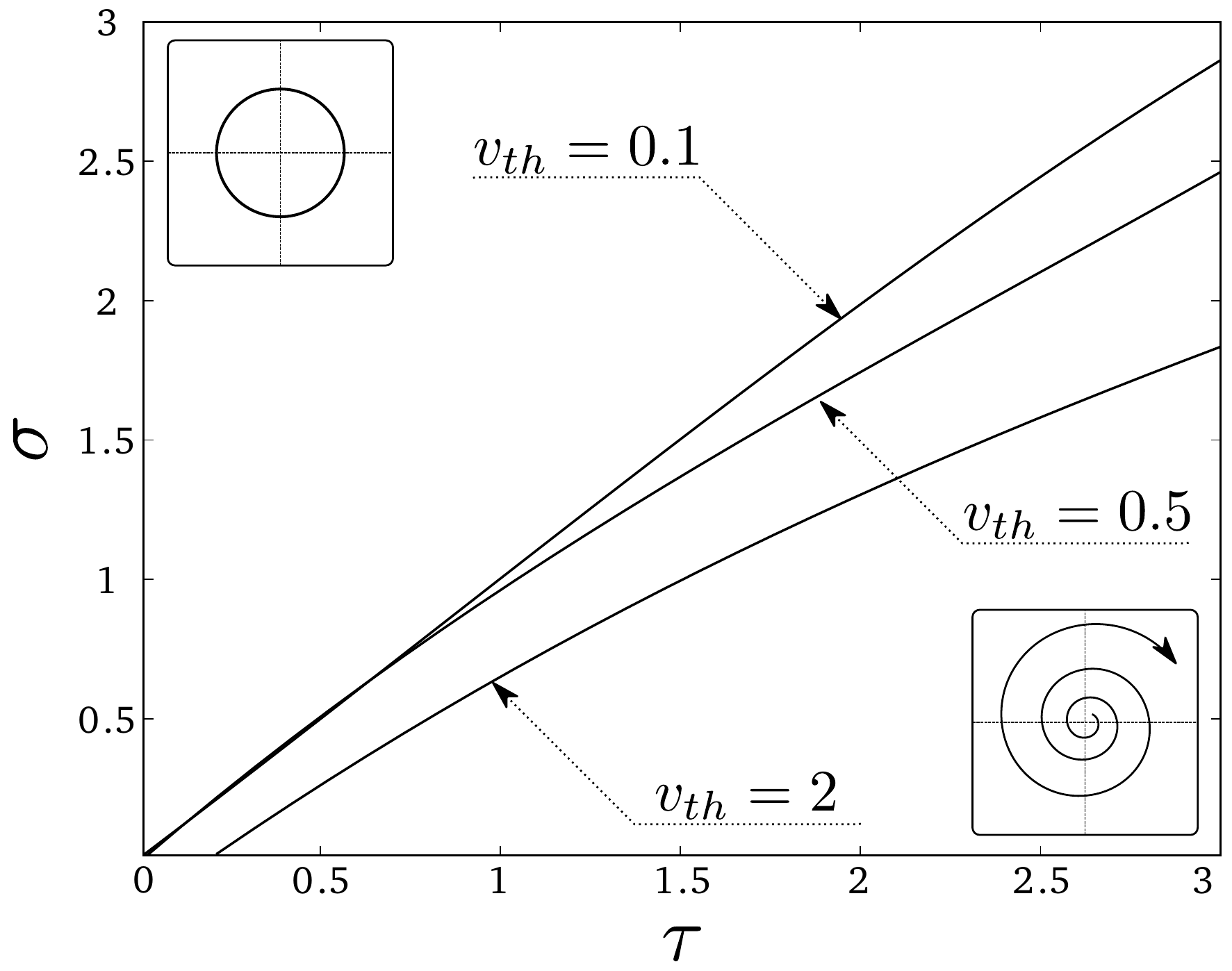}
\caption{Phase diagram of the system~\eqref{eq:model}--\eqref{eq:a_opt}. The curves separating the areas of periodic and unbounded motion were reconstructed numerically for three various values of $v_{\textrm{th}}$. The system motion has been identified as unbounded if the oscillations amplitude had been consistently increasing over large period of time ($10000$ units). }
\label{fig:ph_diag}
\end{figure}

As can be seen from Fig.~\ref{fig:ampl}, there is a boundary in the parameter space $(\tau,\sigma)$ at which the transition from the periodic behavior pattern to the unbounded growth occurs. We found that this boundary depends essentially on the parameter $v_{\textrm{th}}$. Fig.~\ref{fig:ph_diag} illustrates the numerically obtained boundaries for $v_{\textrm{th}}=\{0.1,\,0.5,\,2\}$. One can see that increasing values of the velocity tolerance $v_{th}$ paradoxically lead to a larger area of the parameter space corresponding to the periodic motion around the origin. This may be explained in the following way. The operator characterized by the relatively large response delay $\tau$ may destabilize the system by the delayed and therefore improper actions when trying to act continuously, i.e., to compensate for the tiniest deviations ($v_{\textrm{th}}$ close to zero). However, the operator neglecting small velocity deviations ($v_{\textrm{th}}$ of order unity) can handle maintaining the bounded motion of the 
system, even though by cost of increased motion amplitude. This situation corresponds well to the experimental findings on human control of inverted pendulum, where the continuous control is less efficient in comparing to the discontinuous, or intermittent control~\cite{loram2011human,zgonnikov2014react}. Particularly, in balancing an inverted pendulum human operators who ignore small deviations from the vertical position perform better than the operators trying to react to every detectable deviation~\cite{zgonnikov2014react}.


Finally, we make some remarks on the cases of $\Updelta_v=1$ and $\Updelta_a=1$. When the velocity dynamical trap is absent ($\Updelta_v=1$ or $v_{\textrm{th}}=0$), the system properties described above remain essentially the same, with only minor changes in the form of the limit cycle presented in~Fig.~\ref{fig:phase}. The impact of acceleration dynamical trap is much higher: its absence ($\Updelta_a=1$) makes the operator able to precisely stabilize the system at the origin, instead of just maintaining it in some bounded area (although the operator reaction time $\tau$ should still be relatively small with respect to $\sigma$). %

The following summarizes the discovered dynamical properties of the proposed model:
\begin{itemize}
\item
the operator is basically unable to precisely stabilize the system under control;

\item
well-skilled operator can maintain the oscillations of the system in the vicinity of the desired position;

\item
unintelligent operator fails to control the system, destabilizing it by imprecise actions;

\item
increasing the range of acceptable states may help even the unintelligent operator to get the control over the system.

\end{itemize}

\section{Discussion}
We tackle the problem of mathematical description of human-controlled systems. We build up on the concept of dynamical traps~\cite{lubashevsky2012dynamical},\cite{lubashevsky2003noised},\cite{lubashevsky2002long} that matches the modern para\-digm of discontinuous human control \cite{cabrera2002onoff} and appeals to the existence of a certain region of acceptable states near the desired phase space position. The present paper argues that the dynamical trap notion is more general and extends it to the operators perception of their own actions.

A human operator controlling a dynamical system is usually not capable of selecting or calculating the optimal action strategy that allows to reach and maintain the desired end-state or goal. However, during the control process the operator is able to realize that the currently implemented strategy deviates from the optimal one if this deviation becomes large enough. Once being aware of the mismatch, the operator can adjust the actions until she feels that the current value of control parameter is acceptable. In order to capture this feature of human cognition, we extend the phase space of the dynamical system under human control with the control parameter as an independent phase variable. It enables us to introduce a certain region alongside the optimal strategy in the space of all action strategies; each strategy within this region is treated as acceptable by the operator. The latter region is called the action dynamical trap.

We study an example that describes the behavior of a human operator trying to control a simple dynamical system. The results of the theoretical and numerical analysis of the developed model correspond well to the basic properties of human-controlled systems. Particularly, we elucidate the fact that it is mainly the operator reaction time and capability of suppressing the velocity deviations that determine the system behavior. The system hardly can be precisely stabilized even by the operators with exceptional abilities. Generally, the system is boundedly stable, exhibiting periodic oscillations around the equilibrium or even may be completely destabilized by the actions of operator.

As the problem at hand is concerned with the car following, we feel necessary to note the following. First, the developed model appeals to the notion of oscillator in an extended phase space including acceleration as an independent phase variable. Most of the car-following models, however, are of type~\eqref{eq:general} and consider the driver behavior to be governed by two stimuli: the necessity of keeping a safe headway distance and controlling the relative velocity~\cite{kesting2012traffic}. They operate, in particular, with such notions as the optimal velocity for a given headway and the velocity difference. The linearizion of the corresponding governing equations gives rise to the harmonic oscillator model which can capture a number of fundamental properties of car dynamics~\cite{wagner2012analyzing,wagner2011time,flotterod2013identifiability}. Some of these models explicitly allow for the delay in human reaction  and take the form of delay-differential 
equations. For example, the optimal velocity model with delay~\cite{PhysRevE.58.5429} relates the current car acceleration $a(t))$ with the headway distance $h(t-T)$ and the car velocity $v(t-T)$ taken at the time shifted to the past by the delay time $T$, or, what is the same, $\{h(t),v(t)\} \mapsto a(t+T)$. In this case, expanding $a(t+T)$ into the Taylor series with respect to $T$ within the first order accuracy, i.e., using the approximation 
\[
  a(t+T)= a(t)+T\frac{da}{dt}, 
\]
we can reduce the optimal velocity model with delay to a model operating with the car jerk $da/dt$ as a certain function of the car headway $h$, velocity $v$, and acceleration $a$. It should be noted that exactly this approach was used in establishing the relation between the first order Newell's car-following model and the previous second order model without delay~\cite{kesting2012traffic}. 

Second, there are a number of psychological and action point models taking into account the bounded capacity of drivers in recognizing small variations in the car velocity and headway distance~\cite{brackstone1999car}. They appeal to the human perception thresholds determining the moments when drivers start to correct the motion of their cars. This concept is directly related to the notion of dynamical traps discussed in the present paper.  

The previous studies on the dynamical trap effect have reported on various complex cooperative phenomena in the ensembles of coupled oscillators. Still, the interaction of at least three coupled oscillators was required for the unstable dynamics to emerge without the noise effects. The presented non-Newtonian model demonstrates that even the simple isolated oscillator may exhibit non-trivial behavior in the presence of action dynamical trap.

The model still requires extensive further development. First of all, the stochastic nature of human control has to be taken into account. The conventional approach to this problem, as mentioned previously, is to include the additive or multiplicative Gaussian noise term to the feedback of human operator. Such noise typically substitutes for all the unaccounted minor features of the modeled system, as well as the external random events of small magnitude. However, we feel that this approach may be inappropriate when one wishes to capture the intrinsic stochasticity of human actions. In the present model the corrective actions of an operator are caused by the controlled variable ($a-a_{\textrm{opt}}$) exceeding certain threshold, which is completely deterministic (despite being fuzzy). In other words, the operator reacts exactly to the same value of the system deviation over and over again. We have a strong feeling that the \emph{probabilistic} description of reaction threshold should be developed in order 
to build a more comprehensive model of human control. Another issue is that the control effort generated when the threshold is crossed is again defined in a deterministic way, being a simple linear feedback $\dot a \propto -(a-a_{\textrm{opt}})$. Experimental findings on the stick balancing task reveal that instead of a feedback control human subjects produce open-loop, ballistic-like corrective movements~\cite{loram2002human}. The amplitude of such movements also appears to be probabilistic, which should be given due consideration as well.

Second, we would like to remark that the independence of the two dynamical traps in perceiving the velocity and acceleration deviations is in fact only the zeroth-order approximation. Presumably, the operator internal apparatus recognizing the situations which require active behavior should be regarded as a single mechanism.

Although we mentioned some empirical facts supporting the presented concepts, the action dynamical trap model is still very abstract. It was designed not to mimic a particular real-world system, but rather to highlight that the extended phase space may enable one to capture intricate dynamical properties of the human-controlled systems. The ideas of this proof-of-concept study ought to be developed further into more concrete models of specific human-controlled processes which should be in turn confronted with the experimental data. We believe the phase space extension approach proposed here is rather general and may be employed in a wide class of models where human actions are of primary importance.

\section*{Acknowledgment} 
The work was supported in part by the JSPS ``Grants-in-Aid for Scientific Research'' Program, Grant 24540410-0001.

\end{document}